\documentclass[doublecol]{epl2}
\usepackage{amsmath}
\usepackage{mathrsfs}
\usepackage{amssymb}
\usepackage{amsfonts}
\usepackage{latexsym}
\usepackage{graphicx}
\usepackage{amsfonts,amssymb}
\usepackage{amsmath}
\usepackage{color}
\usepackage{graphics}
\usepackage{array}
\usepackage{fmtcount}
\usepackage{multirow}
\usepackage{enumitem}

\title{Evaluation of node importance in complex networks}

\author{S. Huang \inst{1,2,3,4} \and H.F. Cui \inst{1} \and Y.M. Ding \inst{1,3,4}}

\institute{
    \inst{1} Wuhan institute of Physics and Mathematics, Chinese Academy of Sciences, Wuhan 430071, China\\
    \inst{2} University of Chinese Academy of Sciences, Beijing 100049, China\\
    \inst{3} Key Laboratory of Magnetic Resonance in Biological Systems,Wuhan institute of Physics and Mathematics, Chinese Academy of Sciences, Wuhan 430071, China\\
    \inst{4} National Center for Mathematics and Interdisciplinary Sciences, Chinese Academy of Sciences, Beijing 100190, China
}

\abstract{The assessment of node importance has been a fundamental issue in the research of complex networks. In this paper, we propose to use the Shannon-Parry measure (SPM) to evaluate the importance of a node quantitatively, because SPM is the stationary distribution of the most unprejudiced random walk on the network.  We demonstrate the accuracy and robustness of SPM compared with several popular methods in the Zachary karate club network and three toy networks. We apply SPM to analyze the city importance of China Railways High-speed (CRH) network, and obtain reasonable results. Since SPM can be used effectively in weighted and directed network, we believe  it is a relevant method to identify key nodes in networks.}

\begin{document}
\maketitle

\section{Introduction}
In recent years, complexity research has been a hotspot and frontier of scientific research ~\cite{1,2,3,4,5,6}, and the study of complex networks is an active area inspired largely by the empirical study of real-world networks such as computer networks and social networks, which display substantial non-trivial topological features, with patterns of connection between their elements that are neither purely regular nor purely random. In analyzing the structure organization of a network, identifying important nodes has been a fundamental problem. Node importance can be used in cooperative localization in wireless sensor networks ~\cite{7}, sorting the search results of search engine ~\cite{8}, controlling the spreading of disease ~\cite{9}, preventing blackouts caused by cascading failure ~\cite{10} and so on ~\cite{11,12,13}.

The measures of node importance usually depend either on the local neighborhood or global properties of a network. Examples are degree centrality ~\cite{14} (DC, defined as the degree of a node), betweenness centrality ~\cite{15} (BC, measures the number of times that a shortest path travels through the node), closeness centrality ~\cite{16} (CC, reciprocal of the sum of the length of the geodesic distance to every other node) and eigenvector centrality ~\cite{17,18} (EC, the eigenvector of the largest eigenvalue of an adjacency matrix). There are some other measures based on random walk ~\cite{19} or position attribution such as PageRank ~\cite{20,21} and $k$-shell ~\cite{22} (nodes locate within the core of the network as identified by the $k$-shell decomposition). Generally, a good measure should include information from different scales, both local and global. Some researchers combine these indicators together as a new one to find key nodes in networks ~\cite{23,24}.

In this paper, we would like to use the Shannon-Parry measure (SPM) to evaluate node importance. SPM value of a node is the probability of arriving at that node after a large number of steps, in other words, it is the frequency of a typical long path visit the node. The main idea comes from symbolic dynamics and compatible Markov processes on the network. SPM can characterize the node importance effectively and can be applied to directed networks and weighted networks. Effectiveness of SPM is embodied in its sensibility and robustness. Next Section will provide a complete introduction to our method. In section 3, we compare SPM with some popular methods in the Zachary karate club network ~\cite{25,26} and three toy networks to show the validity of SPM, then apply SPM to find the hub cities of the China Railways High-speed (CRH) network and obtain consistent results. We conclude our work in the last section.

\section{Methods}
We consider a network $G=(V,\ E)$ with a given set of nodes $V$, links $E$ and the number of nodes $N=|V|$. We identify every node $v \in V$ with a natural number $i \leq N$, the network $G$ is represented by its adjacency matrix $A$ with $A_{ij}=1$ if $\{i,\ j\}\in E$, $A_{ij}=0$ if $\{i,j\}\notin E$. If $G$ is an undirected network, $A$ is symmetric, i.e., $A^T=A$. If $A$ is not symmetric, $G$ can be regarded as a directed network. A directed network is called strongly connected if there is a path in each direction between each pair of vertices of the graph.
A directed network corresponds to a subshift of finite type (SFT)--a mathematical object is studied in symbolic dynamics and ergodic theory ~\cite{27}.  Let $Y$ be the set of all infinite admissible sequences of edges, where by admissible it is meant that the sequence is a walk of the graph. Let $T$ be the shift operator on such sequences; it plays the role of the time-evolution operator of the dynamical system. A subshift of finite type is then defined as a pair $(Y, T)$ obtained in this way.  Formally, one may define the sequence of edges as
$$ \Sigma_{A}^{+} = \left\{ (x_0,x_1,\ldots): x_j \in V, A_{x_{j}x_{j+1}}=1, j\in\mathbb{N} \right\}. $$
The shift operator T maps a sequence in the one- or two-sided shift to another by shifting all symbols to the left, i.e.
$\displaystyle(T(x))_{j}=x_{j+1}$. A subshift of finite type is said to be transitive if $G$ is strongly connected. The topological entropy of a SFT is equal to the logarithm of the spectral radius $r$ of the adjacency matrix $A$.

A subshift of finite type may be endowed with different measures, thus leading to a measure-preserving dynamical system ~\cite{27}. A Markov chain is a pair $(P, \Pi)$ consisting of the transition matrix, an $n \times n$ matrix $P=(p_{ij})$ for which all $p_{ij} \ge 0$ and
 $\sum_{j=1}^np_{ij}=1$
for all $i$. The stationary probability vector $\pi=(\pi_i)$ has all $\pi_{i} \ge 0,\sum \pi_i = 1$ and has
$\sum_{i=1}^n \pi_i p_{ij}= \pi_j$.
A Markov chain is said to be compatible with the SFT if $p_{ij} = 0$ whenever $A_{ij} = 0$. The Markov measure of a cylinder set (a path on the network) may then be defined by
    \begin{equation}\label{path}\mu(C[v_0,\ldots,v_s]) = \pi_{v_0} p_{v_0,v_1} \cdots p_{v_{s-1}, v_s}.\end{equation}
The stationary distribution, $\pi=\{\pi_i\}$ can be used to rank the importance of the nodes. In fact, suppose $S_n$ is a typical path with length $n$ generated by the Markov Chain on the graph, denote $S_n(i)$ is the number of times that $S_n$ visit the node $i$, then the frequency of the path visit the vertex $i$ is $S_n(i)/n$, which approaches to $\pi_i$ by ergodic theorem of Markov Chain ~\cite{27}. So a node with larger fraction in the stationary distribution is more important. For undirected network $G$, if we put $P_{ij}=A_{ij}/k_{i}$, where $k_i=\sum_jA_{ij}$ is the degree of node $i$. This means that the walker on node $i$ goes to an adjacent node with the same probability for all neighbors. The stationary distribution of the Markov Chain is $\pi_i= k_i/{\sum_jk_j}$. So the importance of a node is proportional to its degree. This is the rationality of Degree Centrality (DC), which is popular in the rank of nodes.

Note the Kolmogorov-Sinai entropy with relation to the Markov measure is
$s_\mu = -\sum_{i=1}^n \pi_i \sum_{j=1}^n p_{ij} \log p_{ij}$ ~\cite{27}. The variation principle claims that the Kolmogorov-Sinai entropy of any stationary measure of the SFT $(Y, T)$ is not greater than its topological entropy. The stationary measure whose Kolmogorov-Sinai entropy equals to the topological entropy $\log \lambda$ is called an equilibrium state of the SFT ~\cite{27}.

The adjacency matrix $A$ is irreducible if for every pair of nodes $i$ and $j$, there exists a natural number $m$ such that $(A^m)_{ij}$ is not equal to $0$. The adjacency matrix of a strongly connected directed weighted network is irreducible. Fix a node $i$ and define the period of $i$ to be the greatest common divisor of all natural numbers $m$ such that $(A^m)_{ii} > 0$. When $A$ is irreducible, the period of every node is the same and is called the period of $A$. If the period is $1$, $A$ is aperiodic ~\cite{27}.

Let $A$ be an irreducible and aperiodic non-negative $n\times n$ matrix with spectral radius $\rho(A)=\lambda$. By Perron-Frobenius theorem, $\lambda$ is a simple eigenvalue of $A$.
$A$ has a left eigenvector $u$ and a right eigenvector $v$ with eigenvalue $\lambda$ whose components are all positive. Then the $n \times n$ matrix $\{P_{i,j}\}$ defined by
\begin{eqnarray} \label{markov}
P_{ij}=\frac{A_{i,j}v_{j}}{\lambda v_{i}},
\end{eqnarray}
is a transition matrix, which induce a compatible Markov Chain of the SFT $(Y, T)$.  The stationary distribution of the Markov Chain is
\begin{equation} \label{density} \pi_{i}=\frac{u_{i}v_{i}}{\Sigma_{i} u_{i}v_{i}},\end{equation}
and the Kolmogorov-Sinai entropy is $\log \lambda$. The corresponding Markov measure is called the Shannon-Parry measure (SPM) ~\cite{28}, which is the measure of maximal entropy as well as the unique equilibrium state of the SFT $(Y, T)$. In this sense, the compatible Markov Chain is the most unprejudiced Markov Chain (random walk) which can be endowed on the graph $G$. If $A$ is a symmetric matrix, i.e., $G$ is an undirected network, the left and right eigenvectors are coincide, and the detailed balance condition is fulfilled: $\pi_i P_{ij}=\pi_j P_{ji}$. And the measure for a path with length $s$ is independent of intermediate nodes:
 $$\mu(C[v_0,\ldots,v_s]) = \pi_{v_0} p_{v_0,v_1} \cdots p_{v_{s-1}, v_s}=\frac{1}{\lambda^s}\frac{v_j}{v_i}.$$
 Thus all paths having length $s$ and given endpoints $i$ and $j$ are equiprobable ~\cite{29,30}.

In a number of real-world networks, not all edges have the same capacity. In fact, edges are often associated with weights that differentiate them in terms of their strength, intensity, or capacity. For weighted network, the adjacency matrix is not a $0-1$ matrix, $A_{ij}$ is the weight of the edge $\{i, j\}$. Since negative sign of the weight on an edge can be expressed by the direction of the edge, we assume that all of the weights in $G$ are nonnegative, the weighted adjacency matrix $A$ is a nonnegative matrix, which is the mathematical expression of a directed weighted network. Suppose that the weighted adjacency matrix is irreducible and aperiodic, by the Frobenius-Perron theorem of nonnegative matrix, $A$ admits a positive eigenvalue $\lambda$, which is equal to the spectral radius $\rho(A)$, there are left eigenvector $u$ and a right eigenvector $v$ with eigenvalue $\lambda$ whose components are all positive. And the $n \times n$ matrix $\{P_{i,j}\}$ defined by (\ref{markov}) is a transition matrix, the stationary density associated the Markov Chain can also be described by (\ref{density}). As a result, we can use the stationary distribution to rank the nodes in $G$.

The computation of SPM can be realized as follows:
\begin{enumerate}
\item Obtain adjacent matrix $A$ of the strongly connected network, check if $A$ is a non-negative irreducible and aperiodic matrix;
\item Calculate the unique maximal eigenvalue $\lambda$ of matrix $A$, and corresponding left and right eigenvectors $u=(u_{1},\ldots,u_{n})$ and $v=(v_{1},\ldots,v_{n})$.
 Obtain the Markov transition matrix $P$ of the unprejudiced Markov Chain by (\ref{markov});
\item Obtain the Shannon-Parry measure $\pi=\{\pi_1,\cdots, \pi_n\}$ by (\ref{density}),  $\pi_{i}$ is the SPM value of node $i$.
\end{enumerate}

\section{Data and results}
\subsection{Validation of SPM}
First, we use the Zachary karate club network (Fig. 1) as an example to validate the effectiveness of SPM.  Compared with T (total number of times a node gets infected) [26] and other frequently used measures, Table 1 shows the top five nodes according to our measure are 34, 1, 3, 33, 2, which are very close to the results of other methods such as DC, BC and SIS simulation.
\begin{figure}[!htb]
  \begin{center}
  \scalebox{0.5}[0.5]{\includegraphics[width=2.5in]{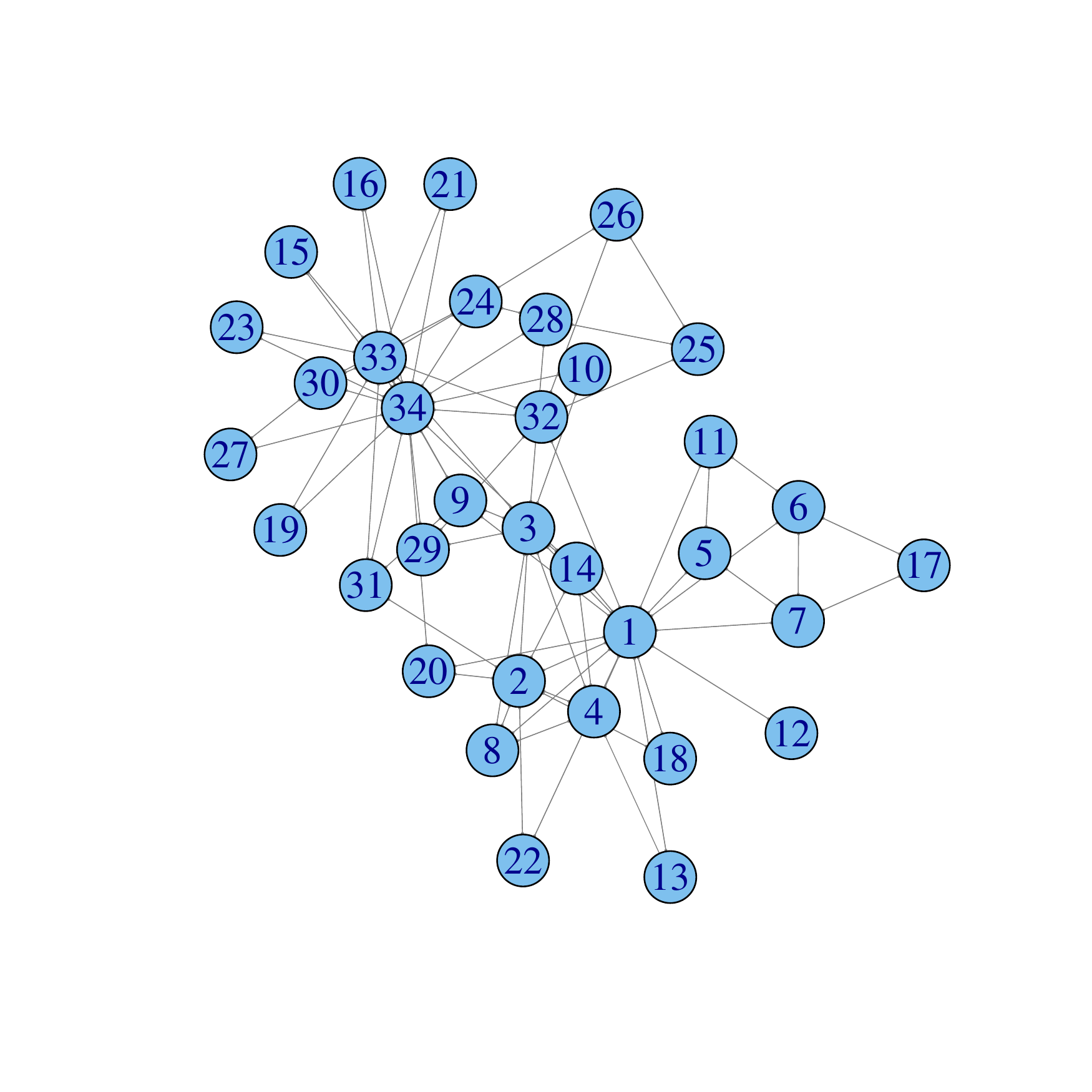}}
  \end{center}
  \caption{\footnotesize The Zachary karate club network ~\cite{25,26}}
\end{figure}

\begin{table}[!htb]
\tabcolsep 0pt \caption{Ranking of the nodes (top 5) in Fig.1 according to SIS simulation (the first and second line) ~\cite{12}, DC, BC and SPM. The top five nodes according to SPM are 34, 1, 3, 33, 2, the result is consistent with other methods.}\vspace*{-20pt}
\begin{center}
\def\temptablewidth{0.49\textwidth}
{\rule{\temptablewidth}{0pt}}
\begin{tabular*}{\temptablewidth}{@{\extracolsep{\fill}} c|c|c|c|c|c}
\hline\hline
{Measure} & \multicolumn{5}{c}{Ranking} \\\hline
$T(\lambda=0.3)$ & 34 & 1 & 33 & 3 & 2 \\\hline
$T(\lambda=0.6)$ & 1 & 34 & 33 & 3 & 2 \\\hline
DC             & 34 & 1 & 33 & 3 & 2 \\\hline
BC             & 1 & 34 & 33 & 3 & 32 \\\hline
SPM             & 34 & 1 & 3 & 33 & 2 \\\hline
\end{tabular*}
{\rule{\temptablewidth}{0pt}}\vspace*{-15pt}
\end{center}
\end{table}

\begin{figure}[!htb]
  \begin{center}
  \includegraphics[width=3in]{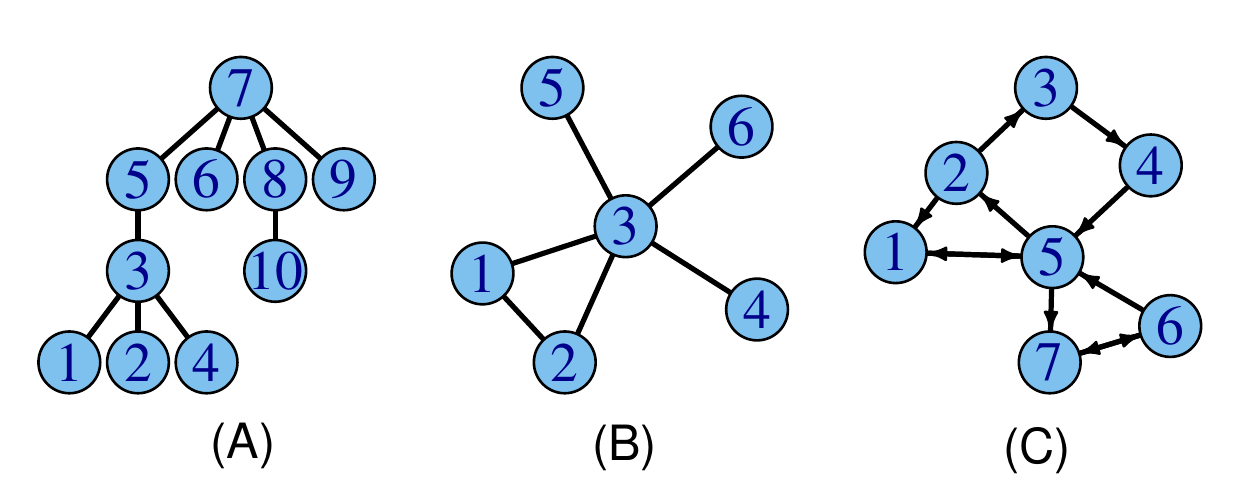}
  \end{center}
  \caption{\footnotesize  Three toy networks. (A) A simple opinion network; (B) A graph with six nodes; (C) A directed graph with seven nodes.}
\end{figure}

\begin{table}[!htb]
\tabcolsep 0pt \caption{Ranking of the ten nodes in Fig. 2 (A) according to DC, $k$-shell and SPM. For each method, the first row represents nodes, the second row is the corresponding node importance. For $k$-shell, the second row represents the step at which the node has been deleted, and the step number is proportional to the node's importance.}\vspace*{-20pt}
\begin{center}
\def\temptablewidth{0.49\textwidth}
{\rule{\temptablewidth}{0pt}}
\begin{tabular*}{\temptablewidth}{@{\extracolsep{\fill}} c|c|c|c|c|c|c|c|c|c|c}
\hline \hline
{Measure} & \multicolumn{10}{c}{Ranking} \\
\hline
\multirow{2}{*}{DC} & 3 & 7 & 5 & 8 & 1 & 2 & 4 & 6 & 9 & 10 \\
\cline{2-11} & 0.22 & 0.22 & 0.11 & 0.11 & 0.06 & 0.06 & 0.06 & 0.06 & 0.06 & 0.06 \\
\hline
\multirow{2}{*}{$k$-shell} & 5 & 7 & 3 & 8 & 1 & 2 & 4 & 6 & 9 & 10 \\
\cline{2-11} & 3 & 3 & 2 & 2 & 1 & 1 & 1 & 1& 1 & 1 \\
\hline
\multirow{2}{*}{SPM} & 7 & 3 & 5 & 8 & 6 & 9 & 1 & 2 & 4 & 10 \\
\cline{2-11} & 0.27 & 0.21 & 0.19 & 0.08 & 0.05 & 0.05 & 0.05 & 0.05 & 0.05 & 0.02 \\
\hline
\end{tabular*}
{\rule{\temptablewidth}{0pt}}\vspace*{-15pt}
\end{center}
\end{table}

\begin{table}[!htb]
\tabcolsep 0pt \caption{Ranking of the six nodes in Fig. 2 (B) according to DC, $k$-shell and SPM. }\vspace*{-20pt}
\begin{center}
\def\temptablewidth{0.49\textwidth}
{\rule{\temptablewidth}{0pt}}
\begin{tabular*}{\temptablewidth}{@{\extracolsep{\fill}} c|c|c|c|c|c|c}
\hline\hline
{Measure} & \multicolumn{6}{c}{Ranking} \\
\hline
\multirow{2}{*}{DC} & 3 & 1 & 2 & 4 & 5 & 6 \\
\cline{2-7} & 0.42 & 0.17 & 0.17 & 0.08 & 0.08 & 0.08 \\
\hline
\multirow{2}{*}{$k$-shell} & 1 & 2 & 3 & 4 & 5 & 6 \\
\cline{2-7} & 2 & 2 & 2 & 1 & 1 & 1 \\
\hline
\multirow{2}{*}{SPM} & 3 & 1 & 2 & 4 & 5 & 6 \\
\cline{2-7} & 0.43 & 0.19 & 0.19 & 0.07 & 0.07 & 0.07 \\
\hline
\end{tabular*}
{\rule{\temptablewidth}{0pt}}\vspace*{-15pt}
\end{center}

\end{table}
\begin{table}[!htb]
\tabcolsep 0pt \caption{Ranking of the seven nodes in Fig. 2 (C) according to SPM.}\vspace*{-20pt}
\begin{center}
\def\temptablewidth{0.49\textwidth}
{\rule{\temptablewidth}{0pt}}
\begin{tabular*}{\temptablewidth}{@{\extracolsep{\fill}} c|c|c|c|c|c|c|c}
\hline\hline
Node   &5       &1      &6      &7      &2      &4      &3   \\\hline
SPM   &0.3208 &0.1790 &0.1577 &0.1577 &0.1059 &0.0394 &0.0394 \\\hline
\end{tabular*}
{\rule{\temptablewidth}{0pt}}\vspace*{-15pt}
\end{center}
\end{table}

Now we show how SPM outperforms other node-importance measures. As can be seen in Fig. 2 (A), it's an undirected and unweighted network consists of ten nodes and nine edges. A moment's inspection ought to suggest that node 3, 5 and 7 are more important than other nodes. Table 2 gives the ranking of the ten nodes according to DC, $k$-shell, SPM. Node 5 and node 8 have the same importance according to DC, but in fact, node 5 locates in a more central position which connects two branches. SPM ranks node 7 and 3 as the top two, and then is node 5, this reconciles the node importance with our expectation since SPM takes into account more factors, and node importance lies not only on the degree, but also the adjacent nodes.

Fig. 2 (B) is a simple network with six nodes. According to $k$-shell (Table 3), node 1,2,3 are equally important, but compared with node 1 and 2, node 3 is much more significant since its favored position. In this case, DC, and SPM characterize the node importance more accurately.

In real world, many networks are directed. In this case, DC and $k$-shell are no longer applicable. In order to show SPM's sensibility, we construct a directed network with seven nodes (Fig. 2 (C)). Table 4 provides a list of nodes ranked in order of decreasing importance.
From the above examples, we can see that SPM is accurate and more sensitive than $k$-shell and DC.

\subsection{Apply SPM in China Railways High-speed (CRH) Network}
To apply SPM in a real complex network, we construct CRH network which consists of commercial train services that have an average speed of 200 km/h (124 mph) or higher. High-speed rail service in China was introduced in 2007, and daily ridership has grown to millions, making the Chinese high-speed rail network the busiest in the world. CRH network is an ideal example of directed and weighted complex network. Identifying traffic hub cities in CRH network is significant, since it can help us to solve practical problems such as traffic arrangement during the "Spring Rush".

Our CRH network contains 39 cities which are selected from three municipalities (Beijing, Tianjin, Shanghai), and two or three main cities in 16 provinces. A directed edge from city $A$ to city $B$ exists if there is a high-speed train trip from $A$ to $B$, and the weight is determined by the number of high-speed train trips per day between them. The data are obtained from the train time table of China Railway Customer Service Center (www.12306.cn) during September 2013, a period which can reflect the normal situation of railway transportation in China. In this way, our CRH network has 39 nodes and 439 edges, as can be seen in Fig. 3 (left panel). The details of the adjacency matrix can be seen in Support Information.  A brief geographical graph is also given in Fig. 3 (right panel), which consists of four main rail lines in China: Beijing-Guangzhou Railway (red line), Beijing¨CShanghai Railway (blue line), Shanghai-Wuhan Railway (yellow line) and Harbin-Dalian Railway (green line).

Since there are more than one 100 high-speed train trips between some cities such as: Beijing--Tianjin, Shanghai--Nanjing, Guangzhou--Shenzhen, and only one high-speed train trip between some cities such as: Beijing--Fuzhou, Wuhan--Jinan.  The weight of Beijing--Tianjin may not be 100 times higher than the weight of Beijing--Fuzhou. We assume that the weight of a given edge has reached saturation if there would be a train running between two cities every other hour on average, so we use the weight $w_{i j}= \min \{ a_{i j}, threshold\}$, where $a_{i j}$ is number of high-speed train trips from city $i$ to city $j$ per day. Since the scheduled running time of high-speed rail in China is supposedly from 6 a.m. to 11 p.m., we suppose the threshold is about 18. According to Table 5, when the threshold is varying from $18$ to $20$, the SPM values of the first ten cities are robust.

\begin{figure}[htb]
  \begin{center}
  \scalebox{1.0}[1.0]{\includegraphics[width=3.2in]{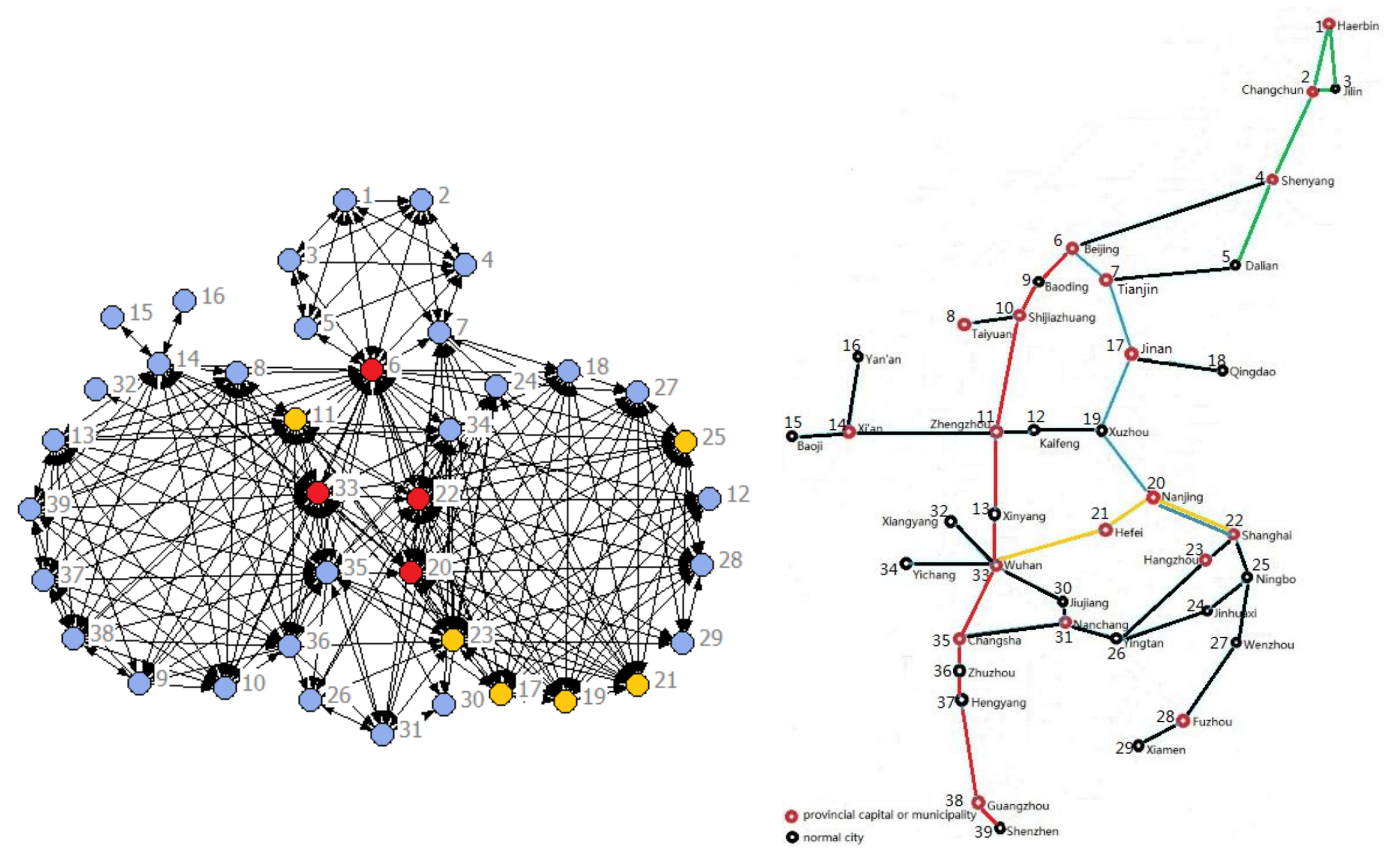}}
  \end{center}
  \caption{\scriptsize (Color online)The left panel is the topological graph of CRH network, it contains 39 nodes and 439 edges, the top 4 cities are marked in red and following 6 cities are marked in yellow. The right panel is a geographical graph of CRH network, the number marked next to each city is the same as in left panel, and the four main high-speed railway lines are marked in red, blue, yellow and green, respectively.}
\end{figure}

\begin{table}[htb]
\tabcolsep 0pt \caption{Ranking of the cities (top 10) according to SPM, the threshold 18, 19, 20 has been listed respectively.}\vspace*{-20pt}
\begin{center}
\def\temptablewidth{0.49\textwidth}
{\rule{\temptablewidth}{0pt}}
\begin{tabular*}{\temptablewidth}{@{\extracolsep{\fill}} c|c|c|c|c|c|c|c|c|c|c}
\hline\hline
\multirow{2}{*}{$18$}  & BJ    & SH    & WH    & NJ    & HZ    & XZ    & ZZ    & JN    & HF   & NB  \\
\cline{2-11}    & 0.109 & 0.101 & 0.100 & 0.095 & 0.064 & 0.057 & 0.054 & 0.052 & 0.041 & 0.038  \\
\hline
\multirow{2}{*}{$19$}  & BJ    & SH    & WH    & NJ    & HZ    & XZ    & ZZ    & JN    & HF   & NB  \\
\cline{2-11}    & 0.110 & 0.102 & 0.101 & 0.097 & 0.064 & 0.057 & 0.054 & 0.053 & 0.041 & 0.038  \\
\hline
\multirow{2}{*}{$20$}  & BJ    & SH    & WH    & NJ    & HZ    & XZ    & ZZ    & JN    & HF   & NB  \\
\cline{2-11}    & 0.110 & 0.103 & 0.100 & 0.099 & 0.064 & 0.057 & 0.054 & 0.053 & 0.041 & 0.039  \\
\hline
\end{tabular*}
{\rule{\temptablewidth}{0pt}}\vspace*{-15pt}
\end{center}
\end{table}

From Table 5, we can see that Beijing, Shanghai, Wuhan, and Nanjing play most important roles in the CRH network, these top 4 cities account for nearly 40 percent of SPM value. Beijing is the Capital of China, and it's a transportation hub for two CRH railway lines: the red line and the blue line. Shanghai is an international metropolis which connects east China to Beijing and central China. Wuhan lies in the middle of the red line and runs high-speed railway to Shanghai, it's favorable geographical position makes it a transportation hub connected north China and south China. Nanjing is a fast developed city lies in the blue line which connects Shanghai to Beijing. Jinan and Xuzhou also rank well because they are located in the key positions of the blue line and are connected with Zhengzhou of the red line. In addition to geographical position, the economic situation also plays a role in being CRH hub. Compared with existing knowledge about China's most important railway transportation hubs, our results reflect the reality well.

\section{Conclusions}
We have proposed a node-importance method, the Shannon-Parry measure (SPM), to quantitatively evaluate the importance of nodes in networks. The main idea comes from symbolic dynamics and ergodic theory of the SFT (subshift of finite type). SPM is not only the stationary distribution of the most unprejudiced compatible Markov Chain on the network \cite{29, 30};
it is also the unique equilibrium state of the corresponding (weighted) SFT \cite{27}; The SPM value of a node is the probability of arriving at that node after a large number of steps. It is also the frequency of a typical long path visit the node. We show the validity of SPM by the Zachary karate club network and three toy networks. SPM admits the following advantages:
\begin{enumerate}
\item SPM incorporates both the local neighborhood and global properties of a network;
\item SPM can characterize the node importance effectively, and can be applied to directed networks and weighted networks;
\item SPM is sensitive and robust.
\end{enumerate}
We applied SPM to identify the hub cities of the China Railways High-speed (CRH) network and obtained rational results. We think that SPM  is a relevant method to identify key nodes in complex networks.
\section{Acknowledgment}   This work was partially supported by National Basic Research Program of China (973 Program) Grant No.2011CB707802, 2013CB910200 and National Science Foundation of China Grant No.11201466.

\end{document}